\documentclass[12pt]{article}

\input epsf
\def\t{\widetilde}
\def\beq{\begin{eqnarray}}
\def\eeq{\end{eqnarray}}

\def\mpl{M_{\rm Pl}}
\def\e{{\epsilon}}

\def\n{{\cal N}}

\def\lsim{\mathrel{\rlap{\lower3pt\hbox{\hskip0pt$\sim$}}
     \raise1pt\hbox{$<$}}}         
\def\gsim{\mathrel{\rlap{\lower4pt\hbox{\hskip1pt$\sim$}}
     \raise1pt\hbox{$>$}}}         

\begin{document}


\begin{flushright}
{ NYU-TH-04/03/23}
\end{flushright}
\vskip 0.9cm

\centerline{\Large \bf 
Schwarzschild  Solution}
\vspace{0.5cm}
\centerline{\Large \bf 
In Brane Induced Gravity  }

\vskip 0.7cm
\centerline{\large Gregory Gabadadze and Alberto Iglesias}
\vskip 0.3cm
\centerline{\em Center for Cosmology and Particle Physics}
\centerline{\em Department of Physics, New York University, New York, 
NY, 10003, USA}

\vskip 1.9cm 

\begin{abstract}

The  metric of a Schwarzschild solution in 
brane induced gravity in five dimensions is studied. 
We find a nonperturbative solution for which 
an exact expression on  the brane is obtained.  
We also  find a linearized solution in the bulk and argue that a 
nonsingular exact solution in the entire space should exist. 
The exact solution on the brane is highly nontrivial 
as it interpolates between different distance scales.
This part of the metric is enough to deduce 
an important  property --  the ADM mass of the solution is 
suppressed compared  to the bare mass of a static source. 
This screening of the mass is due to nonlinear interactions  which 
give rise to  a nonzero curvature  outside the source.
The curvature extends away from the source to a certain 
macroscopic distance that coincides with the would-be strong 
interaction scale. The very same curvature shields the source 
from strong coupling effects. 
The four dimensional law of gravity, including the correct tensorial 
structure, is recovered at observable distances. 
We find  that the solution has no vDVZ discontinuity
and show that the  gravitational field on the brane is always weak, 
in spite of the fact that the solution is nonperturbative.

\newpage

\end{abstract}

\section{Introduction, Discussions, and Summary}

Exact static solutions in models of gravity carry a great deal of 
information on the gravitational theories themselves. Hence, finding 
these solutions in models that modify gravity at large distances is an 
important and interesting task. In the present work we will study the 
Schwarzschild solution  in the DGP model \cite {DGP} 
where gravitational interactions are modified at large cosmological 
distances. It is complicated to find this solution  
since even at scales much larger than the Schwarzschild radius 
of a source, full nonlinear treatement is required 
\cite {DDGV}.  The first approximate 
solution was obtained in Ref. \cite {Andrei} and subsequently by 
the authors of  Refs. \cite {MassimoBH,Lue,Sio,Nicolis}. 
The solution should interpolate between very different distance scales. 
These scales are: the 4D gravitational radius of the source 
of a mass $M$, $r_M\equiv 2G_N M$,
the large distance crossover scale $r_c\sim 10^{28}$ cm, and an 
intermediate scale, first discovered by Vainshtein in massive 
gravity \cite {Arkady}, which in the DGP model reads as follows \cite{DDGV}:
\beq
r_*\,\equiv\,\left (r_M\,r_c^2  \right )^{1/3}\,.
\label{int}
\eeq
This is a scale at which nonlinear interactions in a naive 
perturbative expansion in $G_N$ become comparable with the linear terms 
(we will discuss below the physical meaning of this scale in detail). 
For a source such as the Sun, the hierarchy of the scales is as follows:
\beq
r_M\,  \ll \,r_*\, \ll r_c\,. 
\label{hierarchy}
\eeq
In most of the work, unless stated otherwise, we will be 
discussing sources that are smaller than $r_*$. In Refs. 
\cite {DGP,DDGV,Andrei,MassimoBH,Lue,Sio,Nicolis} the approximate 
solutions for such sources were found in  different 
regions of (\ref {hierarchy}). The main properties of the 
solution can be summarized as follows:

(a) At distances $r\gg r_c$ the 5D Schwarzschield solution
with the 5D ADM mass $M$ is recovered (throughout this work $r$ stands for 
a 4D radius). 

(b) For $r_* \ll r \ll r_c$ the potential scales as in the 
4D Schwarzschild solution.  However, relativistic gravity is a 
tensor-scalar theory that contains the gravitationally 
coupled scalar mode (i.e. the tensorial structure is that of a
5D gravitational theory which contains extra polarizations). 

(c) For $r\ll r_*$  the theory reproduces 
the Schwarzschild solution of 4D General Relativity (GR) 
with a good accuracy.

Perhaps the most important property of the (a-c) solution 
outlined above is the dynamical ``self-shielding'' 
mechanism by which the solution 
protects itself from the would-be strong coupling regime \cite {DDGV}.
Very briefly, the self-shielding can be described as follows: the 
expansion in $G_N$ breaks down at the scale 
$r\sim r_*$ making the perturbative calculations
unreliable below this scale. However, exact nonlinear solutions 
of equations of motion -- which effectively re-sum the series of classical 
nonlinear graphs -- are perfectly sensible well  below the scale $r_*$.
Hence, the correct way of doing the perturbative calculations is first 
to find a classical background solution  of equations of motion 
and then expand around it.

In the present work we attempt to find exactly the  
Schwarzschild solution in the DGP model. We managed
to find explicitly only  a 4D part of the metric. This exact result, 
combined with reasonable boundary conditions in the bulk, 
is sufficient to determine unambiguously a number of crucial properties 
of the solution.  First we confirm the existence of the scale $r_*$ --
this scale enters manifestly our exact solution. 
We also confirm that the self-shielding mechanism outlined 
above takes  place. Furthermore, we obtain deeper insight 
into  the dynamics of the self-shielding, which, 
to the best of our knowledge, has not been emphasized so far 
in the literature: the self-shielding effect takes 
place because a source creates a nonzero scalar curvature 
that extends {\it outside}  the source to a distance 
$r\sim r_*$.  This curvature suppresses nonlinear interactions 
that otherwise would become strong at the scale below $r_*$.
On the other hand, we also find that some of the physical properties 
of our solution differ from those in (a-c). Our solution has the 
following main features:

(A) For $r\gg r_c$, like in (a),  one recovers the 5D 
Schwarzschild solution, however unlike in (a), the 
new solution has the {\it screened} 5D ADM mass
\beq
M_{\rm eff}\, \sim \,M\, \left ({r_M\over r_c}\right )^{1/3}\,. 
\label{screenedmass}
\eeq
The screened mass is suppressed compared to the bare mass 
$M$. Therefore, the new  solution is  energetically favorable 
over the (a-c) solution.

(B) For $ r_*\ll r \ll r_c$ one can think of the solution 
as being a four-dimensional one with an $r$-dependent decreasing 
mass $M(r)\sim r_* r_M /r$. Alternatively, one can simply think of 
the solution just approaching very fast the 5D Schwarzschild metric with 
the screened mass (\ref {screenedmass}), i.e., approaching the 
asymptotic behavior of (A).

(C) For $r\ll r_*$ the results agree with those of (c) with 
a good accuracy. 

The (a-c) and (A-C) solutions both asymptote to Minkowski space
at infinity. However, the way they approach the flat space is different
because of the difference in their 5D ADM masses. 
The (A-C) solution, or any of its parts,  cannot be obtained in 
linearized theory, it is a nonperturbative solution at any distance scale.
Since the mass of the (a-c) solution is  larger than the mass 
of the (A-C) solution, we would expect that the heavier solution 
will eventually  decay into the light one, unless there are some 
topological arguments preventing this decay.
 
The above findings suggest that Minkowski space, 
although globally stable in the DGP model, is {\it locally} 
unstable in the following sense. A static source
placed on an empty brane creates a nonzero 
scalar curvature around it. For a source of the size $\lsim r_*$
this curvature extends to a distance $\sim r_*$. Above this scale 
the solution asymptotes very  quickly to 5D Minkowski space. 
More intuitively, a static source distorts  a brane medium 
around it creating a potential well, and the distortion 
extends to a distance $r\sim r_*$. Since $r_*$ is much bigger than 
the size of the source itself, we can interpret this phenomenon as 
a local-instability of the flat space. 
This local-instability, however, has not been seen in the linearized 
theory \cite{DGP}. It should emerge, therefore, in nonlinear 
interactions and should disappear  when the scale $r_c^{-1}$ is 
taken to zero\footnote{The latter assertion is valid 
since the (A-C) solution, as we will see,  
is regular in the $m_c\to 0$ limit where it turns to 
the conventional 4D solution, i.e., it 
has no vDVZ discontinuity \cite{Veltman,Zakharov}.}.

It is remarkable that the distance scale to 
which the local-instability extends, coincides with the scale
$r_*$ at which the naive perturbative expansion in $G_N$
breaks down.  Therefore, by creating a scalar 
curvature that extends to $r\sim r_*$,
the source shields itself from a would-be strong coupling 
regime that could otherwise appear at distances  
$r\lsim r_*$ \cite {DDGV}:
(i) The coupling of a phenomenologically 
dangerous  extra scalar polarization of a 5D graviton to 4D matter
gets suppressed at distances $r\lsim r_*$ due to the curvature
effects. This is similar to the suppression of the extra 
polarization of a massive graviton on  the AdS background 
\cite{KoganAdS,PorratiAdS}. Indeed, in our case the curvature 
created by the source, although coordinate dependent,  
has the definite sign that coincides with the sign of the 
AdS curvature. As a result, the model approximates 
with a high accuracy the Einstein gravity at $r\ll r_*$ with 
potentially observable small deviations 
\cite {Lunar1,Lue} (see comments below). 
(ii) The self-coupling of the  extra polarizations 
of a graviton, which on a flat background leads to the breakdown of 
a perturbative expansion and to the strong coupling problem, gets now 
suppressed at distances $\lsim r_*$ by the scalar curvature created 
by the source. This is also similar to the suppression
of the self-coupling of the massive graviton polarizations 
on the AdS background \cite {AGS,Luty}. 

The main properties of the classical solution described above 
seem to be universal and should be expected to hold 
in models that modify gravity at large distances. 
To see the viability of this argument  let us look at  the 
4D part of the Einstein equations of the DGP model 
\beq
G_{\mu\nu}\,+\,m_c\,\left [ K_{\mu\nu} - g_{\mu\nu}K \right ]\,
=\,8\pi\,G_N\,T_{\mu\nu}\,,
\label{example}
\eeq
where $G_{\mu\nu}$ is a four-dimensional Einstein tensor,
$T_{\mu\nu}$ is the matter stress-tensor, 
$K_{\mu\nu}(g)$ is a symmetric tensor of 
the brane extrinsic curvature and $K$ denotes its trace
(in other models, e.g., in  massive gravity, 
the extrinsic curvature part will be replaced by the 
mass term). Consider a localized source, a star for instance. 
Outside the source the r.h.s. of (\ref {example}) is zero. 
However, the term $[K_{\mu\nu} - g_{\mu\nu}K]$
need not be zero. This would lead to a nonzero Ricci tensor and scalar. 
This is unlike the Einstein 
gravity where only the Riemann tensor components 
are nonvanishing. The curvature that is produced away from 
the source, however, is small since it is proportional to the 
strength of the source itself multiplied by additional suppression factors 
proportional to powers of $m_c$. 
According to our exact solution the nonzero  curvature extends effectively to 
distances of the order $r_*$, but it is sub-dominant to the 
standard 4D Schwarzschild contribution to the Kretschman scalar, 
except in the region around $r\sim r_*$, 
where the two curvature invariants  are roughly of the same order 
$\sim m^2_c=r_c^{-2}\sim (10^{-42}~{\rm GeV})^2$  
(see Fig. 2 and detailed discussions in section 3). 
One important property of the solution is that 
the gravitational field is weak everywhere  outside the source. 
Nevertheless, the solution is nonperturbative and 
an expansion in $G_N$ 
cannot be used to recover the solution even 
at very large distances $r\gg r_c$.  

In the present paper we are primarily concerned with 
classical sources. Nevertheless, we would like to comment as well
on dynamics of ``quantum'' sources, such as gravitons.
Consider the following academic set-up: 
a toy world in which there is no matter, radiation and/or any classical
sources of gravity -- only gravitons propagate and interact with 
each other in this world. Because of the very same tri-linear 
vertex diagram that leads to the breakdown of the $G_N$ expansion for 
classical sources (see Ref. \cite {DDGV}), the self-interactions of 
the gravitons will become important at lower energy scale than they would 
in the Einstein theory. The corresponding breakdown scale is the 
scale (\ref {int}) adopted to  a quantum source with 
$r_M =1/\mpl$, that is $\Lambda_q^{-1}\sim (r_c^2/\mpl)^{1/3}$ \cite {Luty} 
(see also Ref. \cite {Rubakov} that obtains a somewhat 
different scale). In this set-up the graviton loop diagrams
could in principle generate higher derivative operators that are 
suppressed by the low scale. A theory with such  
high-derivative operators would not be predictive 
at distances below $\Lambda^{-1}_q\sim $ 1000 km or so. 

However, there are two sets  
of arguments suggesting that the above difficulty 
might well be unimportant for the description  of 
a real world which, on top of the gravitons, 
is inhabited by planets, stars, galaxies etc. 
We start with the arguments of Ref. 
\cite {Nicolis}. This work takes a point of view that 
$\Lambda_q$ is a true ultraviolet (UV) cutoff of the theory in a 
sense that at this scale some new quantum gravity degrees of freedom should
be introduced in the model. Nevertheless, as was discussed in 
detail in Ref. \cite {Nicolis},
this should not be dangerous if  one considers a realistic 
setup in which mater  is  introduced into the theory. For instance, 
consider the effect of introducing the classical gravitatonal field 
of the Sun. Because of the gravitational background of the Sun, 
the UV cutoff of the theory becomes a coordinate dependent
quantity $\Lambda_q(x)$. This cutoff grows closer to the source 
where its gravitational field  becomes more and more pronounced,
hence, increasing the value of the effective UV cutoff.
In this approach the authors of Ref. \cite {Nicolis} managed 
to find a minimal required set  of higher dimensional operators that 
are closed with respect to the renormalization group  flow. 
Because of the re-summation of the large classical nonlinear effects 
these operators are effectively suppressed by the coordinate 
dependent scale $\Lambda_q(x)$.  If so, the new UV physics won't 
manifest itself in any measurements \cite {Nicolis}.

Putting all this on a bit more general ground, 
one should {\it define} the model in an external 
background field.  That is, in the action and the partition function 
of the model the metric splits into  two parts $g_{\mu\nu}=
g^{\rm cl}_{\mu\nu} +g^{\rm q}_{\mu\nu}$, 
where $g^{\rm cl}_{\mu\nu}$ stands for the classical background 
metric and $g^{\rm q}_{\mu\nu}$ denotes the quantum fluctuations 
about that metric. The classical part  
satisfies the classical equations of motion with given classical 
gravitational sources such as matter, radiation, planets, stars, 
galaxies, etc...
Then, the effective UV cutoff for quantum fluctuations 
at any given point in space-time 
is a function of the background metric. For a 
realistic setup this effective cutoff is high enough to 
render  the model consistent with observations.

We find the above logic useful and viable. We also think that
the algorithm of Ref. \cite {Nicolis} might be the most convenient 
one for practical  calculations. Nevertheless, there could 
exist deeper dynamical phenomena to the discussions of which we 
turn right now. 
Although our arguments below parallel in a certain  respect 
those of Ref. \cite {Nicolis}, there is a conceptual difference on 
the main issue. Our view, that we will try to substantiate in subsequent 
works, is that the scale $\Lambda_q$ is not a UV scale of the 
model in the sense that some new  quantum gravity degrees of freedom 
should be entering at that scale. We think that all what's needed 
to go above the scale  $\Lambda_q$ is already in the model,
and that this is just a matter of 
technical difficulty of non-perturbative calculations
(or, in other words, is a matter of difficulty of summing up
loop diagrams). The re-summation could in principle cure 
problems at the loop level as well. At this end, we do not see a reason why  
the self-shielding mechanism outlined above  should not be 
operative for ``quantum'' sources too. 
The very same local-instability of Minkowski space 
should manifest itself in nonlinear interactions of quantum 
sources, e.g., gravitons.  The 
local-instability scale in that case is $\Lambda_q$. Hence, we 
would expect that a quantum source creates a curvature around it 
that extends to the distances of the order of $\sim (r_c^2/\mpl)^{1/3}
\sim 1000$ km, and doing so it self-shields itself from the strong 
coupling regime. If this is so, then the problem of loop calculations
boils down to the problem of defining correct variables w.r.t. which 
the perturbative expansion should be performed. In this case the 
field decomposition should take the form:
$g_{\mu\nu}=g^{\rm np}_{\mu\nu} +g^{\rm q}_{\mu\nu}$, 
where $g^{\rm np}_{\mu\nu}$ stands for a nonperturbative 
background metric created by a ``quantum'' source.  
Similar in spirit arguments using a toy model  
were given by Dvali in Ref. \cite {Dvali}.

In this work we solve exactly for the 4D 
part of the Einstein equations of the DGP model. 
Furthermore, we study the bulk metric as far as we can. Here we do 
not have an exact solution. Nevertheless, a number of important and 
reliable properties can be deduced.  
Using the analytic continuation and 
taking advantage of the fact that the bulk asymptotes to Minkowski space,
we argue that a nonsingular solution that matches our 
brane solution should exist in the bulk.  
Furthermore, we find large-distance  asymptotes of the bulk solution.
It is important to point out that irrespective of the form 
of the bulk solution (as long as it is nonsingular),  
we are able to deduce the  properties (A-C) of the system.

Finally, we would like to make  two important comments.
First, the DGP model possesses two branches of solutions 
that are distinguished from each other  
by the bulk boundary conditions. These two branches are 
disconnected.
In this work we concentrate primarily on  the Schwarzschild solution of 
the so-called conventional branch on which the brane and the bulk asymptote 
to Minkowski space at infinity.
However, the second, the so-called ``self-accelerated'' branch \cite {Cedric} 
is extremely interesting as it can be used to describe the 
accelerated expansion of the Universe without introducing dark 
energy \cite {DDG}. In the present work we also find an
exact brane metric for a Schwarzschild source  
on the self-accelerated branch. However, because the asymptotic 
behavior of the solution on this branch is not Minkowski we are not 
able to argue for the existence of a nonsingular bulk solution. 
On the other hand, we do not see any physical reason why this solution 
should not exist in the bulk as well. This branch will be discussed
in detail elsewhere.  Second, it is interesting to note that 
the linearized analysis 
of the DGP model in dimensions six and higher \cite {GGMisha}, 
as well as certain modifications of the five-dimensional 
model \cite {Romb,GG} show no sign of breakdown of perturbation theory
and strong nonlinear effects. It is left for future work to understand
more deeply the interconnections between all these approaches. 

The paper is organized as follow. In Section 2 we set the 
action and equations of motion of the DGP model.
In section 3 we give a qualitative description of the main 
new properties of our solution.
In section 4 we give exact solutions for the metric and extrinsic 
curvature on the brane. 
In section 5 we discuss the absence of the  vDVZ discontinuity
and in section 6 we comment on distinctions between
the $G_N$ and the weak-field expansions in the model. 
In section 7 we use the ADM formalism to 
argue that the solution for the metric and extrinsic curvature can be 
smoothly continued into the bulk space. 
Concluding remarks  are given in Sect. 8.


\section{The setup}

{}We consider the action of the DGP model \cite{DGP}: 
\begin{equation}
 S= M_*^3\int d^5x \sqrt{-g} R+M_P^2\int d^4x\sqrt{-\t g}\t R~.
 \label{action}
\end{equation}
 Here, the $(4+1)$ coordinates are $x^M=(x^\mu,\ y)$, $\mu=0,
\dots, 3$ and $g$ and $R$ are the determinant and curvature of 
the 5 dimensional metric $g_{MN}$,
while $\t g$ and $\t R$ are the determinant and curvatures of the 4 dimensional
metric $\t g_{\mu\nu}=g_{\mu\nu}(x^\mu,y=0)$.

The $(\mu \nu)$ and $(yM)$ equations of motion are respectively,
\begin{eqnarray}
\sqrt{-\t g}\ \t G_{\mu \nu}(x)\delta(y)+{m_c\over 2}\sqrt{-g}\ G_{\mu \nu}
(x,y)&=&0
\label{mn}~,\\
G_{yM}(x,y)& = & 0\label{yM}~, 
\end{eqnarray}
where $m_c=2M_*^3/M_P^2$ is the inverse of the crossover scale
(the Gibbons-Hawking surface term \cite {GH} that guarantees the correct 
Einstein equations (\ref {mn},\ref {yM}) is implied in (\ref {action})).

{}We will study the analog of the Schwarzschild solution in this setup. 
Thus, we consider the most general static metric with spherical symmetry on the
brane and with ${\bf Z}_2$ symmetric line element:
\begin{equation}
ds^2=-{\rm e}^\nu dt^2+{\rm e}^\lambda dr^2+{\rm e}^\mu r^2d\Omega^2+\gamma\ dr
dy+{\rm e}^\sigma dy^2~,\label{ds}
\end{equation}    
where $\nu,\ \lambda,\ \mu,\ \gamma,\ \sigma$ are functions of 
$r=\sqrt{x^\mu  g_{\mu \nu} x^\nu}$ and $y$. The ${\bf Z}_2$ symmetry 
across the brane implies that $\gamma$ is 
an odd function of $y$ while the rest are even.

The jump conditions on the $y$ derivatives of the warp factors across 
the brane (at $y=0$) implied by (\ref{mn}) give\footnote{In
 the second line of (\ref{tt}), (\ref{rr}) and (\ref{thth}) 
the functions are evaluated at $y=0+$.}
\begin{eqnarray}
2{\rm e}^{-\mu+{\lambda\over 2}}&+&2{\rm e}^{-{\lambda\over 2}}\left[-1
+r\lambda_r
-3r\mu_r+{1\over 2}r^2
(\lambda_r-{3\over 2}\mu_r)\mu_r-r^2\mu_{rr}\right] = \nonumber\\ &&
 {m_c r^2\over 
\sqrt{{\rm e}^{\lambda+\sigma}-\gamma^2}}\left((\lambda_r-2\mu_r-
{4\over r})\gamma
-2\gamma_r+{\rm e}^{\lambda}(\lambda_y+2\mu_y)\right) 
~,\label{tt}\\
2{\rm e}^{-\mu+{\lambda\over 2}}&+&2{\rm e}^{-{\lambda\over 2}}\left[-1
-r(\nu_r+\mu_r)-{1\over 2}r^2(\nu_r
+{1\over 2}\mu_r)\mu_r\right]  = \nonumber\\ &&  {m_c r^2\over 
\sqrt{{\rm e}^{\lambda+\sigma}-\gamma^2}}\left(-(\nu_r+2\mu_r+{4\over r})\gamma
+{\rm e}^{\lambda}(\nu_y+2\mu_y)\right) ~,\label{rr}\\
r{\rm e}^{-{\lambda\over 2}}&&
\hskip-1cm\left[\lambda_r-\nu_r-2\mu_r + {1\over 2}r[
(\lambda_r-\nu_r)
(\nu_r+\mu_r)-\mu^2_r]-r(\nu_{rr}+\mu_{rr})\right] =\nonumber\\ &&  
{m_c r^2\over 
\sqrt{{\rm e}^{\lambda+\sigma}-\gamma^2}}
\left( -2\gamma_r -(\nu_r+\mu_r-\lambda_r
+{2\over r})\gamma
+{\rm e}^{\lambda}(\lambda_y+\nu_y+\mu_y)\right)
~,\label{thth}
\end{eqnarray}
corresponding to the $(tt)$, $(rr)$ and $(\theta\theta)$ components.

{}We have not made use of any gauge freedom up to this point. 
A convenient choice is to set $\mu(r, y)=0$ by rescaling 
$r\rightarrow r\ {\rm exp}(-\mu/2)$. Moreover, we set  
$\nu(r,y) =- \lambda(r,y)$ by transformation of the $y$ coordinate.
The resulting line element is (note that this gauge can be 
reached on the brane and 
in the bulk only because we allowed an off-diagonal term in the metric)
\beq
ds^2=-{\rm e}^{-\lambda} dt^2+{\rm e}^{\lambda} dr^2\,+ 
\,r^2d\Omega^2+\gamma\, 
drdy+{\rm e}^\sigma dy^2~.
\label{lineelement1}
\eeq
where $\lambda,\  \gamma,\ \sigma$ are functions of $r$ and $y$.
Our choice is such that the brane is not bent in this coordinate 
system and is located at $y=0$. 
A more conventional diagonal coordinate system 
can be obtained  by a coordinate redefinition after which the interval reads
\begin{equation}
ds^2=-{\rm e}^\nu dt^2+{\rm e}^\lambda dr^2+ r^2d\Omega^2+{\rm e}^
{\beta} dz^2~.
\label{common}
\end{equation}    
Here the functions $\nu $ and $\beta$ are related 
to  $\lambda, \gamma,$ and $ \sigma $. In the $z,r$ coordinate system 
the brane is bent. Typically in brane-world models the 4D part of the Einstein 
equations are not closed. Hence, the induced metric on a brane 
cannot be determined without some input from the bulk equations,
and/or without making certain assumptions about the induced metric itself.
This would also be true in our case. However, in the gauge 
(\ref {lineelement1}), 
we find a subset of the Einstein equations that 
can be closed for the function $\lambda$.  As a result, $\lambda$  can be 
found exactly on the brane. Although the knowledge of $\lambda$ alone is not 
enough to describe the whole gravitational dynamics on the brane (
for instance, this is not enough  for the description  of 5D 
matter geodesics at short distances since transverse derivatives 
of the metric are also entering the 5D geodesic equations) nevertheless, 
combining the knowledge of $\lambda$ with  the 
asymptotic behavior of the other functions in (\ref {lineelement1}) 
that we can also obtain unambiguously, is enough to deduce the 
properties (A-C) of the Introduction. Hence, these properties are ``exact''.

Once a source is placed on the worldvolume,
the brane produces a nonzero extrinsic curvature. 
As a result of this a nonzero 4D intrinsic curvature is also produced. 
We will discuss these issues in detail in the next section.

From (\ref {yM}) we also derive the following equation (in the gauge 
\ref{lineelement1}):
\begin{eqnarray}
0&=& {\rm e}^\sigma ({\rm e}^{\lambda+\sigma}-\gamma^2)
\left(4-4{\rm 
e}^\lambda-8r\lambda_r+2r^2\lambda_r^2-2r^2\lambda_{rr}\right)
\nonumber\\
&&+\,({\rm e}^{\lambda+\sigma}-\gamma^2)\left(4\gamma^2
+\,(4r\gamma-r^2{\rm e}^\lambda\lambda_y-r^2\gamma\lambda_r)\lambda_y
+2r^2\gamma\lambda_{ry}\right)\nonumber\\
&&+\,{\rm e}^\sigma(2\gamma_r-\gamma\sigma_r-\gamma\lambda_r)
(4r\gamma+r^2{\rm e}^\lambda\lambda_y-r^2\gamma\lambda_r)~.
\label{55}
\end{eqnarray}


\section{Structure of the solution}

In the next section we solve the above system of equations 
exactly on the brane and obtain a perturbative solution in the bulk. 
The fact that the exact solution could be found   
on the brane is nontrivial. 
However, the solution  can only be written in an implicit form 
from which the extraction of  useful information requires additional efforts. 
To simplify the reading of the paper we summarize certain nontrivial 
properties of the solution in the present section. The solution itself will 
be derived in the next section.

\subsection{Solution on the brane}

In this subsection we discuss the properties of the solutions 
on the brane, i.e., 
at $y=0$. We find  certain similarities, as well as  
differences, in the 4D part of our solution with  the 
anti-de Sitter-Schwarzschild 
(AdSS) solution of conventional 4D 
General Relativity  (GR) with  a small negative  
cosmological constant $\Lambda$ 

Consider 4D GR with  the cosmological constant $\Lambda =-3m_c^2$. 
Furthermore, 
consider a static source of mass $M$ (a star) and a Schwarzschild radius 
$r_M\equiv  2G_N M$ in this space. In the static coordinate system 
the AdSS solution takes the form
\beq
ds^2 \,=\,-\left (1\,-\,{r_M\over r}\,+\,m_c^2\,r^2\right )\,dt^2 \,+\,
{dr^2 \over \left (1\,-\,{r_M\over r}\,+\,m_c^2\,r^2\right )  }\,+
\,r^2\,d\Omega_2^2\,.
\label{dSS0}
\eeq 
This coordinate system covers the AdSS solution in the interval 
$r_M < r< r_c \equiv m_c^{-1}$.  The following  properties of the AdSS 
solution will be contrasted to our solution.

(i) In the interval $r_M<r<r_c$ there is a new distance scale $r_*$
exhibited by (\ref {dSS0})
\beq
r_*\,\equiv \,\left (r_M\,r_c^2\right )^{1/3}\,. 
\label{rstar}
\eeq
The physical meaning of this scale is as follows.
For $r<r_*$ the Newtonian potential $r_M/r$ in (\ref {dSS0}) 
dominates over the term  $m_c^2r^2$,  while for $r>r_*$ the 
term   $m_c^2r^2$ overcomes  the Newtonian term. Hence,  
$r_* $ is a scale at which the Newtonian and the $m_c^2r^2$ terms 
are equal. This can  also be expressed in terms of invariants. 
Let us define the Kretschman  Scalar (KS)
\beq
R_K\,\equiv\,\sqrt{(R_{\alpha\beta\gamma\delta}^{Sch})^2}\,,
\label{RK}
\eeq 
where $R_{\alpha\beta\gamma\delta}^{Sch}$ is a Riemann tensor of the 
Schwarzschild part of the solution (i.e., of the part that 
survives in the $m_c \to 0$ limit). 
We compare the KS  with the background curvature due to 
the cosmological constant
\beq
\left|R_{\Lambda}\right|\,=\,12\,m_c^2\,.
\label{rmc}
\eeq 
We get
\beq
R_K\,\gg |R_{\Lambda}|\, ~~~~{\rm for}~~~~r\ll r_*\,; ~~~~~
R_K\,\ll |R_{\Lambda}|\, ~~~~{\rm for}~~~~r_*\ll r \ll r_c \,.
\label{curvatures}
\eeq 
Therefore, $r_*$ is a scale at which $R_K \simeq  |R_{\Lambda}| $.
For $r_M \ll r \ll r_*$ the corrections due to the background curvature 
are small and the solution is dominated by the Schwarzschild metric, 
while for $r_*\ll r\ll r_c$ the background curvature terms are bigger that 
the Schwarzschild  terms, both of them still being smaller than 1.

(ii) At $r\gg r_*$ the Schwarzschild part becomes irrelevant 
compared to the AdSS part. 

We will show below that our solution has some of the 
properties described in (i), however, unlike (ii), it 
behaves as 5D Schwarzschild solution at large distances.

The 4D part of our solution (i.e. the solution at $y=0$) 
for $r\ll r_c$ takes the form
\beq
ds^2|_{y=0} = -\left (1- {r_M\over r} + m_c^2 r^2 g(r)\right )dt^2 
+ {dr^2 \over \left (1 -{r_M\over r} + m_c^2 r^2 g(r)\right )  }+
\,r^2\,d\Omega_2^2 .
\label{sol4D}
\eeq 
Like the AdSS solution, the metric (\ref {sol4D})  possesses the $r_*$ 
scale defined in (\ref {rstar}). As we will see below this scale has the 
same physical meaning as in the AdSS case.  For instance,  at $r\ll r_*$ 
\beq
g(r) \,\simeq  \,\left 
({r_*^4\over r^4}\right )^{1\over 1+\sqrt{3}}\,. 
\label{grstar}
\eeq 
Then, it is straightforward to check that 
\beq
{r_M\over r} \gg m_c^2\,r^2 g(r)\,~~~ {\rm for } ~~~ r\ll r_*;~~~~~
{r_M\over r}\, \sim \, m_c^2\,r^2 g(r)\,~~~ {\rm for } ~~~ r\sim r_*\,.
\label{curvsol}
\eeq 
Therefore, the corrections become of the order of the 
${r_M/r}$ term at around $r\sim r_*$.
Moreover, like the AdSS solution, the corrections 
dominate over  ${r_M/r}$ for $r_*\ll r \ll r_c$
turning the 4D behavior of the solution into the 5D 
behavior. The plot of the function is given on Fig. 1.
\begin{figure}
\centerline{\epsfbox{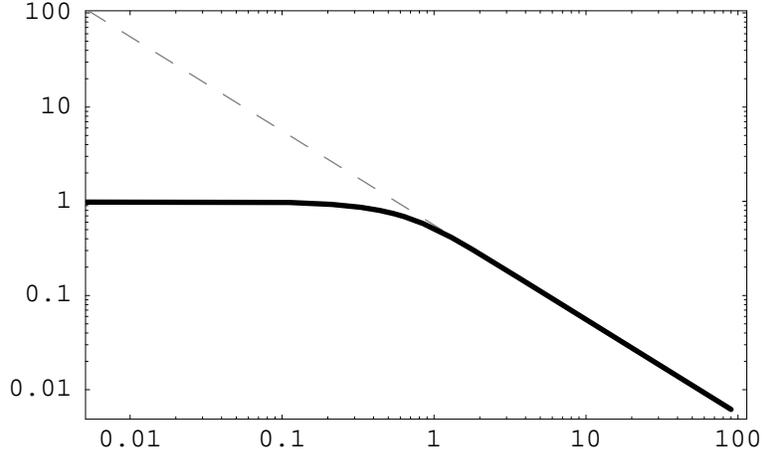}}
\epsfysize=6cm
\vspace{0.1in}
\caption{\small 
The solid line depicts $P(r)/r_M$ (on the 
vertical  axis), wher $P$ is defined 
in (\ref {Pdef}), as a function of $r$ on the horizontal axis. The dashed 
line presents  the function $0.28{r_*}/r$;  The value of 
$r_*$ is set  to 1 on this graph.}
\label{fig1}
\end{figure} 
As in the AdSS case, the corrections to the Schwarzschild solution 
that are proportional to $m_c$ give rise to the four-dimensional Ricci 
curvature $R_{m_c}$. This is interesting since the curvature is completely 
due to the modification of gravity. However, unlike the AdSS case, this 
curvature is not a constant but depends on $r$;  moreover it 
also depends on the strength of the source itself.
The plot of the Ricci curvature is given on Fig. 2.
\begin{figure}
\centerline{\epsfbox{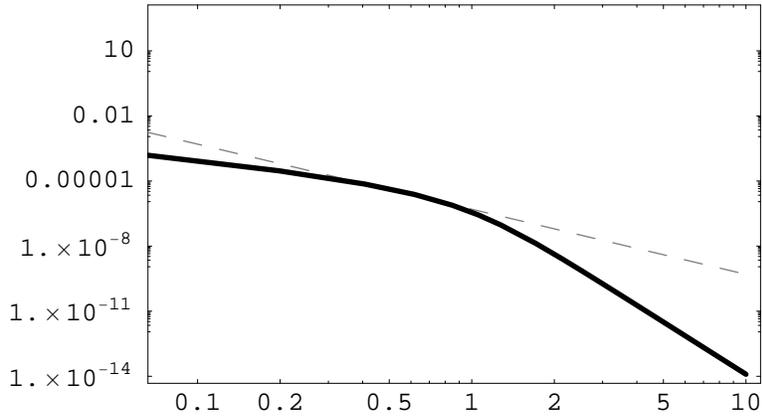}}
\epsfysize=6cm
\vspace{0.1in}
\caption{\small 
The solid line depicts the magnitude of the 
four-dimensional Ricci scalar curvature
(on the vertical axes) as a function of $r$ on the horizontal axes.
The dashed line depicts the dependence of the 4D Kretschmann scalar
on $r$. The value of $r_*$ is set to 1 on this plot.}
\label{fig2}
\end{figure} 

Similar to the AdSS solution the above properties can be expressed 
in terms of the invariants
\beq
R_K\,\gg R_{m_c}~~{\rm for}~~r\ll r_* ~~{\rm and~for}~~r\gg r_*\,;
~~{\rm while}~~ R_K\,\sim  R_{m_c}~~{\rm for}~~r\sim r_*\,.
\label{compcurv}
\eeq 
Unlike in the AdSS solution, however,  
the curvature $R_{m_c}(r)$  decreases very fast 
after $r\gg r_*$. Hence, the induced curvature $R_{m_c}(r)$ is sub-dominant
to $R_K$ everywhere except in the neighborhood of the point $r\sim r_*$
where both of these are of the same order $\sim m_c^2$, see Fig. 2.

Furthermore, unlike the AdSS solution, our 
solution can be presented in the same coordinate system even for $r\ge r_c$. 
This is because there is no horizon at $r=r_c$ and our solution smoothly 
turns into the 5D Schwarzschild solution
\beq
ds^2|_{y=0} \,=\,-\left (1\,-\,{{\tilde r}^2_M\over r^2}\right )\,dt^2 \,
+\, {dr^2 \over \left (1\,-\, {{\tilde r}^2_M\over r^2} \right )  }\,+
\,r^2\,d\Omega_2^2\,.
\label{sol5d}
\eeq 
The key  property of this solution is that the gravitational radius is rescaled
\beq
{\tilde r}_M \,\sim\, r_M\,\left ({r_c\over r_M}     
\right )^{1/3} \,\gg\,r_M\,.
\label{5DrM}
\eeq 
This has an explanation. The gravitational radius grows compared to $r_M$ 
because in the 5D regime the gravitational coupling constant grows. 
However, there is an opposite effect as well. 
In fact, the gravitational radius 
reduces compared to what it would have been in a pure 5D theory with no brane.
This is because the effective mass of the source $M_{\rm eff}$, defined as 
${\tilde r}_M^2 \equiv M_{\rm eff}/M_*^3$, gets screened.  

All the above results could be understood as follows. Consider an empty brane
and an empty bulk. Minkowski space is a solution. Let us localize a 
static source on the brane at $r=0$. The Minkowski solution remains 
globally stable, however, the source, no matter how weak, triggers local 
instability of Minkowski space on a brane in the region $r\le r_*$.
In this patch  Minkowski space is readjusted to a curved space. The 
curvature of the latter depends on the strength of the source, it   
slowly decreases with increasing $r$ but drops fast at  $r> r_*$. 
For an observer at large distances  it looks as if  the  source has  
polarized the medium (brane) around it. This observer measures the 
screened mass (\ref {5DrM}) which also includes the contributions 
of the curvature.

\subsection{Solution in the bulk}

At large enough distances, i.e. $\sqrt{r^2+y^2}\gg r_*$, 
the solution turns into a 5D spherically-symmetric Schwarzschild solution:
\beq
ds^2|_{\sqrt{r^2+y^2}\gg r_*} 
\,\sim\,-\left (1\,-\,{{\tilde r}^2_M\over r^2\,+\,y^2}\right )\,dt^2 \,
+\, {dr^2 \over \left (1\,-\, {{\tilde r}^2_M\over r^2\,+\,y^2} \right )  }\,+
\,(r^2+y^2)\,d\Omega_3^2\,.
\label{bulk5D}
\eeq
However, the 5D spherical symmetric is only an approximation
and does not hold for  $\sqrt{r^2+y^2}\ll r_*$. 
In the latter regime the properties of the solution 
on and off the brane are rather different. The pure 5D spherically-symmetric 
solution (\ref {bulk5D}) 
is squeezed both  on and off the brane but it is more squeezed
on the brane then in the bulk. Hence, the only symmetry of the 
solution that is left is the cylindrical symmetry.


\section{The solutions}

In this section we solve for $\lambda$ on the brane in 
the coordinate system (\ref{lineelement1}).

From the (\ref{tt}) and (\ref{rr}) equations we deduce:
\begin{equation}
\gamma_r={\rm e}^\lambda\lambda_y~,
\end{equation}   
and (\ref{tt}) and (\ref{thth}) can be rewritten as 
\begin{eqnarray}
2{\rm e}^{\lambda\over 2}P_r & = &  {m_c r^2 \over \sqrt{{\rm e}^{\lambda+
\sigma}-\gamma^2}}\left((\lambda_r- {4\over r})\gamma-\gamma_r\right)~,
\label{ttsim}
\\
r{\rm e}^{\lambda\over 2}P_{rr} & = & {2 m_c r^2 \over \sqrt{{\rm e}^{\lambda+
\sigma}-\gamma^2}}\left((\lambda_r- {1\over r})\gamma-\gamma_r\right)~,
\label{ththsim}
\end{eqnarray}
where we have defined 
\beq
P(r)\,\equiv \,r\,(1\,-\,{\rm exp}(-\lambda))\,.
\label{Pdef}
\eeq
Using (\ref{ttsim}) and (\ref{ththsim}) in the $yy$ equation of motion 
(\ref{yM}) we obtain:
\begin{equation}
0= 6m_c^2r^2(2Q+rQ_r)+r^2Q_r^2-8rQQ_r+4Q^2~.\label{yysim} 
\end{equation}
where $Q=P_r$. Let us study this equation. 
First of all we note that the quadratic terms in (\ref {yysim}) 
can only be neglected when $r$ is large. This suggests that 
the naive linearized approximation that neglects the 
quadratic terms is viable only for large distances, 
no matter how weak the source is.  
Hence, for $r\gg m_c^{-1}$ we can neglect the 
last three terms on the r.h.s. of (\ref{yysim}) and then the solution is
\begin{equation}
P={{\tilde r}_M^2\over r}+C_1~,\label{5d}
\end{equation} 
where ${\tilde r}_M$ and $C_1$ are integration constants. 
$C_1=0$ gives the 5D Schwarzschild solution of radius ${\tilde r}_M $. 
In a similar fashion, for 
$r\ll m_c^{-1}$ we can neglect the term 
proportional to $m_c^2$ in (\ref{yysim}). 
The solutions in this case are 
\begin{equation}
P=r_M+C_2 r^{2(2\pm \sqrt{3})+1}~,\label{4d}
\end{equation}
where $r_M$ and $C_2$ are integration constants. In this case, $C_2=0$ 
gives the 4D Schwarzschild solution of radius $r_M$.

{}We will show that there is an interpolating solution between these two 
regimes (regular branch) together with a second solution that becomes 5D 
de Sitter Schwarzschild at large distances (accelerated branch).
In order to find these exact solutions let us rewrite (\ref{yysim}) as 
\begin{equation}
0=U_z^2-4(1+U)U_z-8U(2+U)~, \label{yysim2}
\end{equation}
where $z={\rm ln}(r/r_0)$ and $U=-2{\rm exp}(-2z)Q/3m_c^2r_0^2$ ($r_0$ an 
arbitrary constant).
There are two solutions that are given implicitly by:
\begin{eqnarray}
{\rm ln}\left[-{(1+3U+f)\over U^2(3+3U+\sqrt{3}f)^{2\sqrt{3}}(-5-3U+f)}
\right]&=&8z +C_3~,\label{sol1}\\
{\rm ln}\left[-{(-5-3U+f)(-3-3U-\sqrt{3}f)^{2\sqrt{3}}\over (U+2)^2 
(1+3U+f)}
\right]&=&8z+C_3~,\label{sol2}
\end{eqnarray}
where $f=\sqrt{1+6U+3U^2}$ and $C_3$ an integration constant.

{}Let us first consider the solution (\ref{sol1}). For $U>0$ the left hand side
is a decreasing function of $U$. To obtain the large distance ($r\gg r_*$)
behavior of the solution we consider the limit $U\rightarrow0^+$ ($z\rightarrow
 +\infty$) in which (\ref{sol1}) reads
\begin{equation}
{\rm ln}~U=-4z+C_4+ {\cal O}(U)~,
\end{equation}
that gives (\ref{5d}) upon integration.
The short distance regime ($r\ll r_*$) is found by taking the limit 
$U\rightarrow +\infty$ ($z\rightarrow -\infty$). In this case (\ref{sol1}) is 
given by
\begin{equation}
(1+\sqrt{3}){\rm ln}~U=-4z+C_5+{\cal O}\left({1\over U}\right)~,
\end{equation}
that reproduces the asymptotic behavior of (\ref{4d}) (with the minus 
sign).
Thus, we have found a smooth solution that interpolates between the 4D and 5D 
Schwarzschild solutions on the brane. This corresponds to the regular branch.

{}Let us now study the second solution (\ref{sol2}). For $U<-2$ the left hand 
side of (\ref{sol2}) is an increasing function of $U$. In 
the large distance limit $U\rightarrow -2^-$ ($z\rightarrow +\infty$)
(\ref{sol2}) reads
\begin{equation}
{\rm ln}(-U-2)=-4z+C_4+{\cal O}(U+2)~,
\end{equation}
that gives upon integration
\begin{equation}
P={{\tilde r}_M^2\over r} +m_c^2r^3~,
\end{equation}
where we have absorbed the integration constant $C_4$ in the definition of 
$\tilde r_M$. This is the 5D de Sitter-Schwarzschild solution of the accelerated 
branch. The short distance behavior is obtained by considering 
the $U\rightarrow -\infty$ ($z\rightarrow -\infty$) of (\ref{sol2}).
This gives
\begin{equation}
(1+\sqrt{3}){\rm ln}(-U)=-4z+C_5+{\cal O}\left({1\over U}
\right)~.
\end{equation}
The above equation gives the following behavior of the potentials
for $r_M\ll r\ll r_*$
\beq
\nu\,=\,-\lambda\,\simeq\,-\,{r_M\over r} \,-\,0.84\, m_c^2\,r^2 
\left({r_*\over r}\right)^{2(\sqrt{3}-1)}.
\label{dSS}
\eeq

Below we derive in detail the relation between the mass of the source 
that determines the 4D Schwarzschild radius $r_M$ and the ``screened'' mass 
that  determines the 5D Schwarzschild behavior (with radius $\t r_M$) at large 
distances ($r \gg r_c$). 
In 
order to do that let us rewrite (\ref{sol1}) as
\begin{equation}
F(U)=k u~, \label{a1}
\end{equation}
by defining the variable $u=r^3$, the constant $k={\rm exp}(3C_3/8)/r_0^3$ 
and 
\begin{equation}  
F(U)=\left({(1+3U+f)\over U^2(3+3U+\sqrt{3}f)^{2\sqrt{3}}(5+3U-f)}
\right)^{3/8}~.
\end{equation}
$F(U)$ is a monotonically decreasing function for $U>0$ that diverges in 
the 
limit $U \rightarrow 0+$ and vanishes in the limit $U\rightarrow +\infty$. 
Note that $U=-2P_u/m_c^2$. On the other hand, (\ref{a1}) gives $U=F^{-1}(k 
u)$.
Therefore 
\begin{equation}
-{2\over m_c^2} \int^\infty_0P_u {\rm d}u={1\over k}\int^\infty_0
F^{-1}(k u){\rm d}(k u)~, \label{r_M}
\end{equation}
where the integral on the right hand side is equivalent to the following 
integral
\begin{equation}
\int^\infty_0F(U){\rm d}U\approx 0.43~.
\end{equation}
Imposing the asymptotic behavior of $P$: $P(0)=r_M$ and 
$P(+\infty)=0$ on (\ref{r_M}) gives
\begin{equation}
2kr_M=0.43 m_c^2
\end{equation}
In turn, the large distance behavior of $P$: $P\sim \t r_M^2/r$ obtained 
from 
(\ref{sol1}) gives
\begin{equation}
k (\t r_M r_c)^{3/2}=
\left({3 \over 4\sqrt{2}(3+\sqrt{3})^{\sqrt{3}}}\right)^{3/4}\approx 
0.082~.
\label{tr_M}
\end{equation}
Thus, from (\ref{r_M}) and (\ref{tr_M}) we obtain the exact relation 
between 
the 4D and 5D Schwarzschild radii:
\begin{equation}
r_M=2.6~ \t r_M \sqrt{\t r_M\over r_c}~.
\end{equation}

The 4D Ricci curvature $R_4$ can be readily calculated 
using the expression:
\beq
R_4=-{3\over 2}m_c^2 (U_z+4U)\,.
\eeq
It goes to zero in the large r limit (in the regular branch) and grows 
as one approaches the source at small $r$. Moreover, it gives rise to the 
properties described in the previous section.

In this gauge  we solve for the combination
of $\gamma$ and  $\sigma$ at $y=0^+$ which takes the form
\beq
{\rm e}^{-{\sigma\over 2}}\gamma={\rm arctan}\left({rP_{rr}-4P_r \over 6m_c r}
\right)={\rm arctan}\left({m_cr_0\over 4}{\rm e}^z(2U-U_z)\right)~.
\eeq
The latter expression will be used in the next section to check the 
continuity of the solution in the $m_c\to 0$ limit. 


\section{No vDVZ discontinuity}

The vDVZ discontinuity \cite{Veltman,Zakharov} is an interesting observation 
that the theory in the $m_c\to 0$ limit could differ 
from the one in which $m_c=0$ is set {\it ab~initio}
(i.e., from the Einstein theory). Below we will argue that the 
vDVZ observation is based on  arguments that do not hold 
when the dynamical effects of the mass screening are taken 
into account.

We will show that our solution is continuous in the 
$m_c\to 0$ limit. We present these arguments it two different 
ways. First let us look at the solution of the previous section. 
In the limit $m_c\to 0$ (i.e. $r_c \to \infty $) we find: 
\beq
r_* \to \infty \,;~~~{\tilde r}_M \to \infty\,;~~~ M_{\rm eff}\to 0\,.
\label{limits}
\eeq
Moreover, the off-diagonal term in the solution behaves as follows:
\beq
\gamma e^{-\sigma/2}|_{r\gg r_c}\, \simeq \, {\rm arctan} 
\left  ( { {\tilde r}_M^2 \over m_c r^3} \right )\,\ll 1\,,
\label{offdiag1}
\eeq
which is always small and the regime of the applicability 
of the above expression goes to infinity in the limit $m_c \to 0$. 
Moreover, in the region $r\ll r_*$ we get
\beq
\gamma e^{-\sigma/2}|_{r \ll r_*}\,\to \,{\rm arctan} 
\left  ( {m_c \over m_c^{8 \over 3(1+\sqrt{3})}}\right )\,\to 0\,. 
\label{offdiag2}
\eeq 
Based on the above findings we conclude that the solution turns into 
an exactly 4D Schwarzschild solution in the $m_c \to 0$ limit. 
The regions where it could deviate from the 4D solution, $r\gg {\rm min}
\{ r_*, r_c\}$,  go to infinity.

The vDVZ discontinuity was originally formulated in \cite {Veltman,Zakharov}
in terms of a one-graviton exchange amplitude. It is instructive to see 
the loophole in this formalism as well.
The arguments of \cite {Veltman,Zakharov} go as follows. 
The momentum space amplitude for one graviton exchange between 
the source of a  stress-tensor $T^{\rm source}_{\mu\nu}$ 
and a probe  $T^\prime_{\mu\nu}$ is given by
\beq
{\cal A}_m\,=\, {1\over \mpl^2}\,
{ T^{\rm source}_{\mu\nu}\, T^{\prime \mu\nu}\, -\,{1\over 3}\,
T^{\rm source}\, T^{\prime } \, \over p^2 \,+\,m_c \,p +i\e}\,.
\label{naiveA}
\eeq
(Here $T$'s denote the trace of the stress-tensors.)
The very same amplitude for massless 4D gravity is 
\beq
{\cal A}_0\,=\,{1\over \mpl^2} \,
{ T^{\rm source}_{\mu\nu}\, T^{\prime \mu\nu}\, -\,{1\over 2}
T^{\rm source}\, T^{\prime } \over p^2 + i\e}\,.
\label{Amassless}
\eeq
Hence, in the limit $m_c\to 0$  Eq. (\ref {naiveA}) does not reduce to 
Eq. (\ref {Amassless}).  This is the vDVZ discontinuity.

Our exact solution suggests that these arguments do not hold in an intricate 
way. Let us start with  $r\gg r_*$.  In this 
region the linearized equations turn into  source-free equations 
since the source is only localized  at distances  $r\ll r_*$.
The source-free equations can be solved and the solution 
is a 5D Schwarzschild metric with the Schwarzschild radius being 
a yet unspecified integration constant. This constant should either
be fixed by matching to the solution at $r\le r_*$, 
or by calculating the ADM mass. In either case the curvature 
created by the source will also contribute to the ADM mass.
Therefore, at large distances where the perturbative one-graviton 
exchange is believed to be a good approximation, we have to replace a source 
$T^{\rm source}_{\mu\nu}$  with an effective source  $T^{\rm eff}_{\mu\nu}$
that takes into account the fact that the source distorts the medium
around it. This can be achieved by making a substitution $r_c \to
r_*$ in (\ref {naiveA}). In the case of a static source of mass $M$,
this would replace its mass by an effective 
mass (\ref {screenedmass}). Once this replacement is done, 
we find that the tree level amplitude is discontinuous, however, 
this is not problematic because simultaneously the region in which 
the expression (\ref {naiveA}) is  applicable, i.e. $r\gg r_*$, 
goes to infinity according to (\ref {limits}). 
Hence, no vDVZ discontinuity remains in the theory.

We also note that although the solution is nonperturbative, 
nevertheless the fields in the metric 
remain weak (much smaller than 1) all the way down to distances  
$r\ge r_M $. Hence, the solution 
that we find never exhibits the strong field behavior as long as 
$r\ge r_M $; this is similar to the conventional 4D Schwarzschild 
solution of massless gravity.


\section{Weak-field versus $G_N$-expansion}

Consider small perturbations $h$ about a flat metric $\eta$
\beq
g_{\mu\nu}\, =\,\eta_{\mu\nu}\,+\,h_{\mu\nu}\,.
\label{weakfield}
\eeq
The weak-field expansion (WFE) is a power series expansion 
in $h\ll 1$. In conventional Einstein's theory
the WFE coincides with an expansion in Newton's coupling $G_N$.
However, in the DGP model this is not so because there 
is another dimensionfull parameter in the theory, $m_c$, that 
contaminates the $G_N$-expansion. To see this in some detail let us first look
at the WFE and $G_N$-expansion in Einstein's theory. This can be done 
in the Lagrangian or in equations of motion. We choose the former.
Ignoring the indexes that are not important for our purposes, 
the WFE of the Einstein Lagrangian  reads:
\beq
\mpl^2\,\left (h\,\partial^2\,h\,+\,
h\,\partial \,h \,\partial\, h\,+...\right )
+h\,T\,,
\label{einst}
\eeq
where the dots denote terms that contain only two derivatives but higher 
powers of $h$.  The last term in (\ref {einst}) describes the interactions 
of gravity with matter.
The expansion in the parenthesis of (\ref {einst})
in powers of dimensionless field $h$ is acceptable as long as $h\ll 1$
\footnote{We assume that we are in a regime of applicability of 
the Einstein theory itself, i.e., energy and momenta are smaller  
than $\mpl$}. One the other hand, one could rescale the field
$h\to {\tilde h}/\mpl$. Then the Lagrangian takes the form
\beq
\left ({\tilde h}\,\partial^2\,{\tilde h}\,+\,\sqrt{G_N}\,{\tilde h}\,\partial 
\,{\tilde h}\, \partial \,{\tilde h}\,+...\right )
\,+\,\sqrt{G_N}\, {\tilde h}\,T\,.
\label{einsttilde}
\eeq
The rescaled field ${\tilde h}$ has the canonical dimensionality.
The Newton constant emerges only in the graviton interaction
vertices. Therefore,  one can develop the standard Feynman 
diagram technique as an expansion in powers of $G_N$. 
The results of this expansion would coincide with the results of 
the WFE.

The above arguments do not hold in general in  
theories where gravity gets modified at some large distance scale $r_c$. 
This is because the new dimensionfull parameter 
$r_c=m_c^{-1}$ enters the expansions. 
Below we concentrate on the DGP model to discuss this issue.
The 4D part of the Lagrangian in the DGP model can schematically be 
written  as follows:
\beq
\mpl^2\,\left (h\,\partial^2\,h\,+
\,h\,\partial\, h\, \partial \,h\,+...\right )
\,+\,m_c \,\mpl^2 \,\left (h\,  \partial \,h \,+\, h\, h\, \partial \,
h \, +\,...\right )\, +\, h\,T\,.
\label{einstDGP}
\eeq
Let as look at the two terms in (\ref {einstDGP}). The cubic and higher 
powers in the first 
parenthesis can be neglected w.r.t. the quadratic terms as long as $h\ll 1$.
Likewise,  the cubic and higher powers in the second  parenthesis
can be neglected w.r.t. the quadratic terms as long as $h\ll 1$. 
However, the cubic term in the first parenthesis cannot 
be neglected w.r.t. the quadratic terms in the second parenthesis unless
the derivative of the filed is very small $\partial h\ll m_c$. 
To see why this is important one should look at the 
trace equation (since we dropped  the indices in all 
the expressions above we have not made a distinction 
between the traceless and trace Einstein equations).
The trace equation is subtle because 
the linearized bulk equations make the coefficient of the first 
quadratic term in (\ref {einstDGP}) vanish identically
and the nonlinear term is the leading one \cite {GG}.
Therefore, at short distances $\partial h \gg m_c$ some of 
the nonlinear terms in (\ref {einstDGP}) cannot be neglected. This 
is despite of the fact that the fields are weak ($h\ll 1$)!

Let us now turn to the $G_N$-expansion of (\ref {einstDGP}).
It is clear that the rescaling $h\to {\tilde h}/\mpl$ does not lead to a 
Lagrangian with a single dimensionfull coupling as in (\ref {einsttilde}).
Instead we get the nonlinear vertices that contain  $G_N$ as well as $m_c$.
This leads to dramatic consequences.  Certain nonlinear but tree-level
Feynman diagrams contain inverse powers of $m_c$ and diverge in the 
$m_c \to 0$ limit \cite {DDGV}.  For finite $m_c$ the same diagrams 
give rise to  the breakdown of the $G_N$-expansion below the 
scale $r_*$ \cite {DDGV}. However, this is an artificial difficulty that
is brought about by the expansion in $G_N$. As we have seen in the previous 
sections the exact solution is regular in the $m_c\to 0$ limit and 
fields are weak as log as $r\gg r_M$. The metric  is 
non-analytic in $G_N$ showing that the $G_N$-expansion is not valid 
even in the regions where it naively would be expected to work. 
Nevertheless, as we discussed in the first section, 
the problems with the $G_N$ expansion can be fixed. Classically this is 
achieved by summing up the nonlinear corrections
\cite {DDGV}. In the quantum theory things are more subtle 
since the summation of the diagrams is hard to perform while
the expansion in $G_N$  can lead to the appearance of 
certain higher-dimensional operators that are  suppressed by 
unacceptably low scale \cite {Luty}. However, as was shown in 
\cite {Nicolis} and discussed in section 1, even in this case one 
can formulate a quantum perturbation theory in $G_N$ in which the 
counter-terms eliminate the dangerous loop-induced operators and 
the $G_N$-expansion remains a useful tool.

Note that a slight modification of the model can give rise to 
a theory in which the $G_N$-expansion  does not break down 
below $r_*$ \cite {GG}. In this case the calculations can be performed 
straightforwardly as in the Einstein theory. It remains to be see if 
the approach of Ref. \cite {GG} can be implemented in a full nonlinear 
theory.


\section{Existence of the bulk solution}

In the previous sections we presented an exact solution for the metric and 
extrinsic curvature on the brane. Below this will  be used to argue that the 
solutions can be smoothly continued into the whole bulk.  
For this we use the ADM formalism \cite {ADM}.

We introduce the {\it lapse} scalar field  ${\cal N}$, and 
the {\it shift} vector field $\n_\mu $ according to the standard 
rules:
\beq
g_{\mu y}\,\equiv \, \n_\mu\,,~~~g_{yy}\,\equiv \,\n^2\,+\,
g_{\mu\nu}\,\n^\mu \,\n^\nu\,. 
\label{adm}
\eeq 
After integration by parts the action (\ref {action}) takes the form:
\beq 
S= M_*^3\int d^4x dy \sqrt{-{\rm det} g_{\mu\nu}} 
\,\n\,\left (\t R \,+\,K^2 -K_{\mu\nu}K^{\mu \nu} \right )
\,+ \, M_P^2\int d^4x\, \sqrt{-\t g}\t R~,
\label{ADMaction}
\eeq
where $K=g^{\mu \nu}K_{\mu \nu}$ is the trace of the extrinsic curvature 
\beq
K_{\mu \nu}\,=\,{1\over \n}\,\left (\partial_y g_{\mu \nu} -D_\mu \n_\nu
-D_\nu \n_\mu \right )\,,
\label{K}
\eeq
and $D_\mu $ is a covariant derivative with the metric $g_{\mu \nu}$.
Note that the Gibbons-Hawking term implied in (\ref {action})
is canceled in (\ref {ADMaction}) after integration by parts.

What we found in the previous section are the quantities:
\beq
g_{\mu \nu}(x,y=0)\,=\,{\t g}_{\mu \nu}(x)\,~~~~{\rm and }~~~~ 
K_{\mu \nu}(x,y=0)\,. 
\label{initialdata}
\eeq 
This data on a brane can be considered as an 
initial data from which the metric and extrinsic curvature 
in the bulk could be reconstructed.
Let us look at the bulk equations of motion that 
follow from (\ref {ADMaction}) by taking variations w.r.t.
$\n$, $\n_\mu$ and $g^{\mu\nu}$ we get respectively
\begin{eqnarray}
\t R\,& = & \,K^2 - K_{\mu \nu}^2 \,,
\label{1} \\
D_\mu K\,&=& \,D^\nu K_{\mu \nu} \,,
\label{2}\\
g_{\lambda  \mu} g_{\beta \nu}{1\over d}\partial_y (d (Kg^{\lambda \beta}-
K^{\lambda \beta}))
 \,&=&\, -\n G_{\mu \nu}+{1\over 2}g_{\mu \nu}\n (K^2 - K_{\beta \lambda }^2) 
+D_\mu D_\nu \n  \nonumber \\ &&
 \hskip -5.2cm -g_{\mu \nu}D^2 \n  +2D_{[\lambda}(K^\lambda_{\nu]} \n_\mu)
+g_{\mu \nu} D^\lambda (K_{\nu\lambda})  
 - D^\lambda (\n_\lambda K_{\mu \nu})+ 2\n K^\lambda_{[\mu} K_{\nu]\lambda}\,,
\label{3}
\end{eqnarray}
where $d\equiv \sqrt{-{\rm det} g_{\mu \nu}} $ and $G_{\mu \nu}$ 
denotes the Einstein tensor. We look at (\ref {1} -\ref{3}) as 
at a system  of differential equations in the $y$ variable. 
Then, Eq. (\ref {1}) is just an algebraic equation 
since it contains no $y$ derivatives except those that 
are already reabsorbed into the definition of $K$. The same is true
for Eq. (\ref {2}). Furthermore, it is not difficult to check that 
(\ref{1}) is satisfied at $y=0+$ if (\ref{ttsim})-(\ref{yysim}) are
fulfilled. Moreover, Eq. (\ref{2}), when considered 
as an equation defining $\gamma_y(r,y=0+)$,
is satisfied by our solution at $y=0+$.
Hence, the only true differential equation which evolves the 
initial data into the bulk is (\ref {3}) that 
contains first derivative w.r.t. $y$ on its l.h.s.

Given the initial value formulation (\ref{1})-(\ref{3}) local existence of
the
bulk solution is guaranteed~\cite{Wald,Anderson:2003ge}. The problem of global
existence of the bulk solution, {\it i.e.} the geodesic completeness of
the bulk solution obtained by the evolution equation (\ref{3}), is 
not easy to establish in general.  However, in the present context
we can take an advantage of the fact that the bulk metric
asymptotes to Minkowski space. Indeed, a number  of theorems exist 
for asymptotically flat spaces (see \cite{Book}). In particular, in
\cite{Book} the evolution along a timelike direction of data given on a 3 
dimensional surface is shown  to be smooth and geodesically complete under 
the assumption of
{\it strong} asymptotic flatness and a smallness condition on the initial data.
The smallness condition is:
\beq
{\rm
sup}\left((d_0^2+1)^3R_{\mu\nu}^2\right)+\int\sum_{l=0}^{3}(d_0^2+1)^{l+1}
\nabla^lK^2+\int\sum_{l=0}^{1}(d_0^2+1)^{l+3}\nabla^l B^2<\infty\,,
\label{small}
\eeq
where $d_0$ is the geodesic distance from a point $o$ in the 
initial data surface and   
$B$ is the curl of the traceless part of $R_{\mu\nu}$. Our initial data is
strongly asymptotically flat since
\begin{eqnarray}
\t g_{\mu \nu}&=&(1+{\tilde r_M^2\over r^2})\delta_{\mu \nu}+
{\cal O}(r^{-{5\over2}})\,,\nonumber\\
K_{\mu \nu}&=&{\cal O}(r^{-{7\over2}})\,, \nonumber
\end{eqnarray} 
for sufficiently large $r$,  and satisfies the smallness condition as well.
This suggests
that the solution can be extended into a smooth, geodesically complete and 
asymptotically flat bulk, however, \underline{does not} 
constitute a full proof.

We also mention that one could easily find a linearized form of the solution
at $r\gg r_c$. This solution coincides with the linearized 5D Schwarzschild 
metric  in which the integration constant is fixed to the screened mass,
i.e., ${\tilde r}_M^2 = r_*\,r_M$.


\section{Concluding remarks}

In this section we discuss certain interesting issues 
that arise as a byproduct of our studies  
and for which further detailed work is needed. 

\vspace{0.1cm}

{\it (i) Geodesic motion of matter and light}. It is interesting to look at 
the geodesic motion of matter and light in the metric  (\ref {lineelement1}).
The geodesic equation 
for the light rays propagating on the brane is $dr/dt = e^\lambda $.
The useful  form of $\lambda $ is given in Eq. (\ref {sol4D}). From 
this we conclude that for $r\lsim r_*$ the light propagates on the brane 
as it would  in conventional 4D Einstein gravity with some 
corrections that are negligible everywhere except in the region $r\sim r_*$.
As we discussed, for a solar mass object this distance is at $10^{20}$
cm and the above effect will be overshadowed by gravitational 
effects of other sources. 
 
The matter geodesics are more complicated however. 
The 5D geodesic equation on the brane  contains 
derivatives of $\gamma $ across the brane that multiply $dy$.
Since the $\gamma_y $ is singular on the brane, it can give rise 
to finite contributions even though it is multiplied by the vanishing
differentials. If so, the motion of 5D matter will get additional contributions
from the extrinsic curvature of the brane. 
Also we could not manage  to solve  the matter geodesic equations, 
it looks like that the extrinsic curvature 
part is canceling, according to (\ref {example}),  
the modifications of 4D gravity that appear in  
$\lambda$. On the other hand, 4D geodesics for localized matter in 4D are 
determined by the 4D induced metric governed by $\lambda$.  

\vspace{0.1cm}

{\it (ii) Comments on Black Holes}. In the present  
work we were dealing with a macroscopic source on the brane, a star 
for instance. In the context of the DGP model 
these type of sources are assumed to be localized on a brane by a 
certain mechanism not related to gravity itself. If the sources  
were not localized, then, as is known \cite {Marco},
the brane with the induced graviton kinetic term has effectively repulsive 
gravity and it would push any source off the brane. For instance, 
ordinary black holes cannot be held on the brane. However, charged 
black holes could still be quasi-localized if the corresponding gauge 
fields are 
localized. It would be interesting to see how the properties of charged 
black holes would differ. It is also very interesting to 
understand in detail the structure of the 
horizon of a black hole in the bulk. Our preliminary 
findings suggest that it should have a cylindrical form, where 
the cylinder extends from a brane into the bulk to a distance that is
bigger that ${\tilde r}_M$ but smaller than $r_*$. Further detailed 
investigations are needed to understand the validity and 
implications of these observations.

%

\section*{Acknowledgments}

{} We would like to thank Jose Blanco-Pillado, Cedric Deffayet, Gia Dvali, 
Andrei Gruzinov, Arthur Lue, Rob Myers, Roman Scoccimarro, Misha
Shifman, Glenn Starkman and Matias Zaldarriaga for useful discussions. 
The work of AI is supported  by funds provided by New York University.

\end{document}